\documentclass[preprint,showpacs,preprintnumbers,amsmath,amssymb]{revtex4}
\input psfig.sty
\usepackage{graphicx}% Include figurthebibliographye files
\usepackage{dcolumn}% Align table columns on decimal point
\usepackage{bm}% bold math
 
\begin{document}
 
\def\be{\begin{equation}}
\def\ee{\end{equation}}
 
%\preprint{APS/123-QED}

\title{Relativistic statistical theory and generalized \emph{stosszahlansatz}}

%%%%%%%%%%%%%%%%%%%%%%%%%%%%%%%%%
\author{R. Silva} \email{rsilva@on.br,rsilva@uern.br}
\affiliation{Observat\'orio Nacional, Rua Gal. Jos\'e Cristino 77,
20921-400 Rio de Janeiro - RJ, Brazil}

\affiliation{Universidade do
Estado do Rio Grande do Norte, 59610-210, Mossor\'o, RN, Brazil}

%\affiliation{Universidade do
%Estado do Rio Grande do Norte, 59610-210, Mossor\'o, RN, Brasil}

\date{\today}% It is always \today, today,
 
\begin{abstract}
We have investigated the proof of the $H$ theorem within a manifestly covariant approach by considering the relativistic statistical theory developed in [Phy. Rev. E {\bf 66}, 056125, 2002; {\it ibid.} {\bf 72}, 036108 2005]. In our analysis, however, we have not considered the so-called deformed mathematics as did in the above reference. As it happens
in the nonrelativistic limit, the molecular chaos hypothesis is slightly
extended within the $\kappa$-formalism, and the second law of thermodynamics
implies that the $\kappa$ parameter lies on the interval [-1,1]. It is shown
that the collisional equilibrium states (null entropy source term) are described
by a $\kappa$ power law generalization of the exponential Juttner distribution,
e.g., $f(x,p)\propto (\sqrt{1+ \kappa^2\theta^2}+\kappa\theta)^{1/\kappa}\equiv\exp_\kappa\theta$, with
$\theta=\alpha(x)+\beta_\mu p^\mu$, where $\alpha(x)$ is a scalar, $\beta_\mu$
is a four-vector, and $p^\mu$ is the four-momentum. As a simple example, we
calculate the relativistic $\kappa$ power law for a dilute charged gas under the
action of an electromagnetic field $F^{\mu\nu}$. All standard results are readly recovered in the particular limit $\kappa\rightarrow 0$.
\end{abstract}
 
\pacs{05.90.+m;05.20.-y;05.45.+b}% PACS,
                              %display SL99desired
\maketitle
 
%\section{introduction}

\emph{Introduction.} The Boltzmann's famous $H$ theorem, which guarantees positive-definite entropy
production outside equilibrium, also describes the increase in the entropy of an
ideal gas in an irreversible process, by considering the Boltzmann equation.
Roughly speaking, this seminal theorem implies that in the equilibrium
thermodynamic the distribution function of an ideal gas evolves irreversibly
towards maxwellian equilibrium distribution \cite{TO}. In the special
relativistic domain, the very fisrt derivation of this theorem was done by
Marrot \cite{marrot46} and, in the local form, by Ehlers \cite{ehlers61}, Tauber and
Weinberg \cite{tauber61} and Chernikov \cite{chernikov63}. As well known, the $H$
theorem furnishes the Juttner distribution function for a relativistic gas in
equilibrium, which contains the number density, the temperature, and the local
four-momentum as free parameters \cite{juttner}.
 
Recently, this theorem has also been investigated in the context of a nonextensive statistic mechanics (NSM) \cite{ana92}. In fact, the NSM has been proposed as
a possible extension of the classical one, being a framework based on the
deviations of Boltzmann-Gibbs-Shannon entropic measure \cite{T95b}. It is worth mentioning
that most of the experimental evidence supporting a NSM are related to the
power-law distribution associated with the classical $N$-body problem
\cite{SPL98}. More recently, based on similar arguments, Abe
\cite{abe97,abe2004} and Kaniadakis \cite{k1,k2} have also proposed other
entropic formulas. In this latter ones, the $\kappa$-entropy emerges in the context of
the special relativity and in the so-called {\it kinetic interaction principle}
(KIP). In particular, the relativistic $H$ theorem in this approach has also been investigated through a self-consistent relativistic statistical theory \cite{kaniadakis05} and through the framework of nonlinear kinetics \cite{KaniaH01}.

Actually, this $\kappa$-framework leads to a class of one parameter deformed
structures with
interesting mathematical properties \cite{kaniad2001}. A
connection with the generalized Smoluchowski equation was
investigated \cite{chava04}, and a fundamental test, i.e., the so-called
Lesche stability was also checked in the $\kappa$-framework
\cite{kaniadakis04}. It was also shown that it is possible to obtain a consistent
form
for the entropy (linked with a two-parameter deformations of logarithm
function), which generalizes the Tsallis, Abe and Kaniadakis logarithm
behaviours \cite{kania05}. In the experimental viewpoint, there exist some
evidence
related with the $\kappa$-statistic, namely, cosmic rays flux,
rain events in meteorology \cite{kaniad2001}, quark-gluon plasma
\cite{miller03}, kinetic models describing a gas of interacting atoms and
photons \cite{rossani04}, fracture
propagation phenomena \cite{cravero04}, and income distribution \cite{drag03},
as well as construct financial models \cite{bolduc05}. In the theoretical front,
some studies on the canonical quantization of a classical system has also been
investigated \cite{scarfone05}.
 
From the mathematical viewpoint, the $\kappa$-framework is based on
$\kappa$-exponential and $\kappa$-logarithm functions, which is defined by
\begin{equation}\label{expk}
\exp_{\kappa}(f)=
(\sqrt{1+{\kappa}^2f^2} + {\kappa}f)^{1/{\kappa}},
\end{equation}
\begin{equation}\label{expk1}
\ln_{\kappa}(f)=
({f^{\kappa}-f^{-\kappa})/2\kappa}.
\end{equation}
The $\kappa$-entropy associated with $\kappa$-framework is given by 
\begin{equation}\label{e1}
S_{\kappa}(f)=-\int d^{3}p f
[a_{\kappa}f^{\kappa}+a_{-\kappa}f^{-\kappa}+b_{\kappa}]
\end{equation}
which recovers standard Boltzmann-Gibbs entropy $S=-\int f \ln f d^3 p$ in the
limit $\kappa\rightarrow 0$; see Ref. \cite{k1,k2} for details. Hereafter the Boltzmann constant is set equal to unity for the sake of simplicity.

Previous works have already discussed some specific choices for the constants
$a_\kappa$ and $b_\kappa$, i.e., for the pair [$a_\kappa = 1/2\kappa$,
$b_\kappa=0$], the Kaniadakis entropy reads \cite{k1,abe2004}
\begin{equation}
S_\kappa = - \int d^3 p f\ln_\kappa f =  - \langle{\ln_\kappa (f)\rangle}.
\end{equation}
The other choices are given by : [$a_\kappa= 1/2\kappa(1+\kappa)$,
$b_\kappa=0$] \cite{k1}, [$a_\kappa= 1/2\kappa(1+\kappa)$,
$b_\kappa=-a_\kappa-a_{-\kappa}$] \cite{abe2004} and [$a_\kappa=
Z^\kappa/2\kappa(1+\kappa)$, $b_\kappa=0$] \cite{k1}. In
this latter one, the $\kappa$-entropy is given by
\cite{k1,kaniad2001}
\begin{equation} \label{tsaent}
S_{\kappa} = - \int{d^{3}p
            \left(\frac{z^{\kappa}}{2\kappa(1+\kappa)}f^{1+\kappa}
                 -\frac{z^{-\kappa}}{2\kappa(1-\kappa)}f^{1-\kappa}\right)}.
 \end{equation}

In this letter, we intend to extend the nonrelativistic $H$ theorem within the
Kaniadakis framework to the special
relativistic domain through a mannifestly covariant approach. As we shall see, our approach does not consider the so-called deformed mathematics \cite{k2}. Rather, we show a proof for the $H$ theorem based on similar arguments of Refs. \cite{silva05,lima01}, e.g., a generalization of the molecular chaos hypothesis and of the four-entropy flux.

\vspace{0.2cm}

\emph{Classical $H$ Theorem.} We first recall the basis for the proof of the standard
$H$ theorem within the special relativity. As well known, the $H$ theorem is
also based on the molecular chaos
hypothesis (Stosszahlansatz), i.e., the assumption that any two
colliding particles are uncorrelated. This means that the two
point correlation function of the colliding particles can be
factorized
\begin{equation}\label{Boltzmann2}
 f(x, p, p_1) = f(x, p) f(x, p_1),
\end{equation}
or, equivalently,
\begin{equation}\label{Boltzmann2a}
 \ln f(x, p, p_1) = \ln f(x, p) + \ln f(x, p_1),
\end{equation}
where $p$ and $p_1$ are the four-momenta just
before collision and the particles have four-momentum $p\equiv p^\mu=(E/c,\bf
{p})$
in each point $x \equiv x^\mu=(c t,\bf{r})$ of the space-time,
with their energy satisfying $E/c=\sqrt{{\bf p}^2+m^2 c^2}$. In order to complete the proof of the $H$ theorem, we combine the Boltzmann equation with the four-divergence of the four-entropy flow, i.e.,
\begin{equation}
S^{\mu}=-c^2 \int {d^{3}p\over E} p^{\mu} f \ln f. 
\end{equation} 

In this concern, it is possible to show that the relativistic
$\kappa$ entropy is consistent with a slight
departing from ``Stosszahlansatz". Basically, this means the replacement of the logarithm
functions appearing in
(\ref{Boltzmann2a}) by $\kappa$-logarithmic (power laws) defined by Eq. (\ref{expk1}). In
reality, it is worth mentioning that the validity of the chaos molecular
hypothesis still remains as a very controversial
issue \cite{Zeh}.

\emph{Generalized $H$ Theorem.} In order to investigate $H$ theorem in the context of the $\kappa$ statistics, we first consider a relativistic rarified
gas containing $N$ point particles of mass $m$ enclosed in a volume $V$, under
the action of an external four-force field $F^\mu$. Naturally,
the states of the gas must be characterized by a Lorentz invariant one-particle
distribution function $f(x,p)$, which the quantity $f(x,p) d^3xd^3p$ gives, at
each time $t$, the number of particles in the volume element $d^3xd^3p$
around the particles space-time position $x$ and momentum ${\bf
p}$. By considering that every influence of a power law statistic must happen within the collisional term of Boltzmann equation (see also \cite{silva05, lima01}), we assume that the temporal evolution of the
relativistic distribution function $f(x,p)$ is given by the
following $\kappa$-transport equation
\begin{equation}\label{relBot}
p^{\mu}\partial_\mu f+m F^\mu{\partial{f}\over\partial
p^{\mu}}=C_{\kappa}(f),
\end{equation}
where $\mu = 0,1,2,3$ and $\partial_\mu=(c^{-1}\partial_t,\nabla)$ indicates
differentiation with respect to time-space coordinates and
$C_\kappa$ denotes the relativistic $\kappa$-collisional term.  Following  the
same physical arguments concerning the collisional term from approach of Refs. \cite{silva05, lima01}, we have that  $C_\kappa(f)$ has the general form
\begin{equation}
C_\kappa(f) = {c\over 2} \int F \sigma R_\kappa (f,f'){d^3 p_1\over E_1}
d\Omega,
\end{equation}
where $d\Omega$ is an element of the solid angle, the scalar $F$ is the invariant
flux, which is equal to $F=\sqrt{(p_\mu p^\mu_1)^2 -m^4 c^4}$, and $\sigma$ is the differential cross section of the collision $p + p_1\rightarrow p'
+ p'_1$; see Ref. \cite{degrrot} for details. All quantities are defined in the center-of-mass system of the colliding particles. Next, we observe that $C_\kappa$ must
be consistent
with the energy, momentum, and the particle number conservation
laws, and its specific structure must be such that the standard
result is recovered in the limit $\kappa\rightarrow 0$.
 
Here,  the $\kappa$-generalized form of molecular chaos hypothesis is also a difference of
two correlation functions
\begin{eqnarray*} \label{eq:2.12}
R_{\kappa} (f,f') = \exp_{\kappa}\left(\ln_{\kappa}{ f'}+
                          \ln_{\kappa}{
                          f'_{1}}\right)
\end{eqnarray*}
\begin{equation} -
            \exp_{\kappa}\left(\ln_{\kappa}{ f} +
                          \ln_{\kappa}{f_{1}}\right)
,
\end{equation}
where primes refer to the distribution function after collision, and $\exp_\kappa
(f)$, $\ln_\kappa(f)$, are defined by Eqs. (1) and (2). Note that for 
$\kappa= 0$, the above expression reduces to
$R_0=f'{f'}_1-ff_1$, which is exactly the standard molecular chaos hypothesis. In the present framework, the $\kappa$ four-entropy
flux reads
\begin{equation}
S^{\mu}_\kappa = - c^2\int{{d^{3}p\over E}
            p^\mu\left(\frac{z^{\kappa}}{2\kappa(1+\kappa)}f^{1+\kappa}
                 -\frac{z^{-\kappa}}{2\kappa(1-\kappa)}f^{1-\kappa}\right)},
\end{equation}
where for $\mu = 0$, the quantity $c^{-1}S^0_\kappa$ stands for the local Kaniadakis` entropy density, as given by (\ref{tsaent}).
Indeed, by taking the four-divergence of $S^\mu_\kappa$, i.e.,
\begin{equation}
\partial_\mu S^{\mu}_\kappa \equiv \tau_\kappa= -c^{2} \int{d^3p\over E} p^\mu\partial_\mu f\ln_\kappa f,
\end{equation}
and combining with $\kappa$ relativistic Boltzmann equation (\ref{relBot}), we obtain the
source term
\begin{equation}
\tau_\kappa = -{c^3\over 2} \int{d^3p\over E}{d^3p_1\over E_1}d\Omega F \sigma R_\kappa\ln_\kappa f. 
\end{equation}
Next, $\tau_\kappa$ can be written in a more
symmetrical form by using some elementary symmetry operations, 
which also take into account the inverse collisions. Let us 
notice that by interchanging $p$ and $p_1$ the value of the
integral above is not modified. This happens because the scattering cross
section and the magnitude of the flux are invariants
\cite{degrrot}. The value of $\tau_\kappa$ is not
altered if we integrate with respect to the variables $p'$ and
$p'_1$. Although changing the sign of $R_\kappa$ in this step
(inverse collision), the quantity ${d^3pd^3p_1/p^0 p^0_1}$ is also
a collisional invariant \cite{degrrot}. As one may check, such considerations imply that the
$\kappa$-entropy source term can be written as
\begin{eqnarray*}  \label{eq: 2.26}
\tau_{\kappa}(x)={c^3\over 8} \int {d^3p\over E}{d^3p_1\over E_1}d\Omega F \sigma (\ln_{\kappa}{f'_{1}} +\ln_{\kappa}{f'}-
\end{eqnarray*}
\begin{eqnarray*}
-\ln_{\kappa}{f_{1}} -\ln_{\kappa}{f})[\exp_{\kappa}(\ln_{\kappa}{
f'}+\ln_{\kappa}{f'_{1}})-
\end{eqnarray*}
\begin{eqnarray}
-\exp_{\kappa}(\ln_{\kappa}{f} +
\ln_{\kappa}{f_{1}})].
\end{eqnarray}
Now, by using the expression relating $\kappa$ and the speed of light $c$ (see Eq. (8.8) in Ref. \cite{k2}), the $\kappa$-entropy source results in
\begin{eqnarray*}\label{eq: 2.27}
\tau_{\kappa}(x)= l_\kappa\int {d^3p\over E}{d^3p_1\over E_1}d\Omega F \sigma (\ln_{\kappa}{f'_{1}} +\ln_{\kappa}{f'}-\ln_{\kappa}{f_{1}} -\ln_{\kappa}{f})
\end{eqnarray*}
\begin{eqnarray}
\times [\exp_{\kappa}(\ln_{\kappa}{
f'}+\ln_{\kappa}{f'_{1}})-\exp_{\kappa}(\ln_{\kappa}{f} +
\ln_{\kappa}{f_{1}})],
\end{eqnarray}
where $l_\kappa$ is given by
\begin{equation}\label{lk}
l_\kappa = {1\over 8} \left({T\over m}\right)^{3/2}\left({1-\kappa^2\over \kappa^2}\right)^{3/4}.
\end{equation}
As is well known, the irreversibility thermodynamics emerging from molecular collisions is quite obtained if $\tau_\kappa(x)$ is positive definite. This condition is guaranted only when:
i) The integrand in (\ref{eq: 2.27}) is not negative, and ii) $l_\kappa$ is positive. Indeed, the expressions of the integrand
%\begin{subequations}
\begin{equation}
(\ln_{\kappa}f'+\ln_{\kappa} {f}'_1-\ln_{\kappa}
f-\ln_{\kappa} f_1)
\end{equation}
and
\begin{eqnarray}
\exp_{\kappa}(\ln_{\kappa}f'+\ln_{\kappa}{f'}_1)
-\exp_{\kappa}(\ln_{\kappa}f+\ln_{\kappa}f_1)
\end{eqnarray}
%\end{subequations}
is always positive for any pair of distributions $(f,f_1)$ and $(f',f'_1)$. This means that the sign of the four-entropy source is determined by the sign  of $l_\kappa$. Therefore, the positiveness of $\tau_\kappa$ must be mantained \cite{degrrot,GP} only when the deformation parameter $\kappa$ appearing in $l_\kappa$ belong to range [-1,1]. It is also  worth emphasing that this range for $\kappa$ was obtained from completely different physical arguments in Refs. \cite{k2,kaniadakis05}.

For the sake of completeness, let us derive the version of the Juttner distribution within the $\kappa$-statistic. Such a expression is the relativistic version of the $\kappa$-distribution \cite{kaniad2001}, and must be obtained as a natural consequence of the relativistic $H$ theorem. At this point, it is interesting to emphasize that such a distribution already has been introduced by Kaniadakis through a variational problem in a selfconsistent approach; see Ref. \cite{kaniadakis05} for details. 
The $H$ theorem states that $\tau_\kappa=0$ is a necessary and sufficient
condition for equilibrium. Since the integrand of
(\ref{eq: 2.26}) cannot be negative, this occurs if and only if
\begin{equation}
\ln_{\kappa} f'+\ln_{\kappa} {f}'_1=\ln_{\kappa}f+\ln_{\kappa} f_1,
\end{equation}
where the four-momenta are connected through a conservation law $$p^\mu + p^\mu_1 = {p'}^\mu+{p'}^\mu_1,$$ which is valid for any binary collision. Therefore, the above sum of $\kappa$-logarithms
remains constant during a collision: it is a summational
invariant. In the relativistic case, the most general collisional 
invariant is a linear combination of a constant plus the
four-momentum $p^\mu$; see Ref. \cite{degrrot}. Consequently, we must have
\begin{equation}\label{joia}
\ln_{\kappa} f (x,p)=\theta=\alpha(x)+\beta_\mu p^\mu,
\end{equation}
where $\alpha(x)$ is a scalar, $\beta_\mu$ a four-vector, and $p^\mu$
is the four-momentum. By using that $\exp_\kappa (\ln_\kappa f) = f$, we may rewrite
(\ref{joia}) as 
\begin{equation} \label{eq:2.34a}
f(x,p)=\exp_{\kappa}\theta=(\sqrt{1+ \kappa^2\theta^2}+\kappa\theta)^{1/\kappa},
\end{equation}
with arbitrary space and time-dependent parameters $\alpha(x)$ and
$\beta_\mu(x)$. Some considerations on the function $f(x,p)$ are given as follows. First, this is the most general expression which leads to a vanishing
collision term and entropy production, and reduces to Juttner
distribution in the limit ${\kappa \rightarrow 0} $. However, it is not
true in general that $f(x,p)$ is a solution of the transport
equation. This happens only if $f(x,p)$ also makes the left-hand-side
of the transport equation (\ref{relBot}) to be identically null.
Nevertheless, since (\ref{eq:2.34a}) is a power law, the
transport equation implies that the parameters $\alpha(x)$ and
$\beta_\mu (x)$ must only satisfy the constraint equation
\begin{equation}
p^\mu\partial_\mu \alpha(x)+p^\mu p^\nu\partial_\mu
\beta_\nu(x)+m\beta_\mu(x) F^\mu(x,p)=0.
\end{equation}
Second, the above expression is also the relativistic version
of the $\kappa$ distribution \cite{k1}. Here, this was obtained through the different approach from the one used in Refs. \cite{k2,kaniadakis05}. 

As an example, let us now  consider a relativistic gas
under the action of the Lorentz 4-force $$F^\mu(x,p)=-(q/mc) F^{\mu\nu}(x)p_\nu,$$
where $q$ is the charge of the particles and $F^{\mu\nu}$ is the
Maxwell electromagnetic tensor. Following standard lines, it is
easy to show that the local equilibrium function in the presence of an
external electromagnetic field reads
\begin{equation}
f(x,p)=\exp_{\kappa}\left[{\mu-[p^\mu+c^{-1}q A^\mu (x)]U_\mu\over
T}\right],
\end{equation}
where $U_\mu$ is the mean four-velocity of the gas, $T(x)$ is the
temperature field, $\mu$ is the Gibbs function per particles, and
$A^\mu(x)$ the four potential. As physically expected, note also that the above expression reduces, in
the limit $\kappa \rightarrow 0$, to the well known expression
\cite{degrrot,Hakim}
\begin{equation}
f(x,p)=\exp\left({\mu-[p^\mu+c^{-1}q A^\mu (x)]U_\mu\over 
T}\right).
\end{equation}

\emph{Conclusions.} In Refs. \cite{silva05,lima01} we have discussed the $H$ theorem in the context from the Kaniadakis and Tsallis statistic within the nonrelativistic and relativistic domain. Based on the generalization of the chaos molecular hypothesis and the entropic measure, it was shown the proof of the $H$ theorem in both domain. In this letter, by considering the same arguments on the chaos molecular and entropy, and regardless of the deformed mathematics introduced in Ref. \cite{k2}, we have studied a $\kappa$-generalization of the relativistic Boltzmann's kinetic equation along the lines defined by the $\kappa$-statistic. As in Refs. \cite{k2,kaniadakis05}, which obtain constraints on $\kappa$, e.g. [-1,1] (by considering conections between deformed mathematics and special relativity), here we have shown that the second law of thermodynamics also corroborated this constraint, i.e., that the deformation parameter $\kappa$ is restricted in the range above. In reality, since the basic results were obtained through a manifestly covariant way, their generalization to the general relativistic domain may be readly derived. 

Finally, we also emphasize that our proof is consistent with the standard laws describing the microscopic dynamics, and reduce to the familiar Boltzmann proof in the limit $\kappa\rightarrow0$.

\noindent {\bf Acknowledgments:} The author is grateful to J. S. Alcaniz for helpful discussions and a critical reading of the manuscript. This work was supported by Conselho Nacional de Desenvolvimento Cient\'{i}fico e Tecnol\'{o}gico - CNPq (Brasil).

\end{document}